\documentclass[aps,prb,twocolumn,english,superscriptaddress,citeautoscript,preprintnumbers,amsmath,amssymb,floatfix,footinbib]{revtex4-2}
\usepackage[breaklinks=true,colorlinks,citecolor=blue,linkcolor=blue,urlcolor=blue]{hyperref}
\usepackage{amsmath}
\usepackage{graphicx}
\usepackage{dcolumn}
\usepackage{float}
\usepackage[utf8]{inputenc}
\usepackage{color}

\begin{document}
	\title{Topological Surface States and Anisotropic Magnetotransport in SnSb$_{6}$Te$_{10}$}
	\author{Mohit Mudgal}
	\affiliation{Department of Physics, Indian Institute of Technology Kanpur, Kanpur 208016, India}
	\affiliation{Research Institute for Synchrotron Radiation Science (HiSOR), Hiroshima University, Higashi-Hiroshima 739-0046, Japan}
    \author{Subhajit Mandal}
	\affiliation{Department of Physics, Indian Institute of Technology Bombay, Mumbai 400076, India}
    \author{Bishal Das}
	\affiliation{Department of Physics, Indian Institute of Technology Bombay, Mumbai 400076, India}
    \author{Priyanka Meena}
	\affiliation{Department of Physics, Indian Institute of Technology Kanpur, Kanpur 208016, India}
    \affiliation{Department of Molecular Chemistry and Materials Science, Weizmann Institute of Science, Israel}
    \author{Amarjyoti Choudhury}
	\affiliation{Department of Physics, Indian Institute of Technology Roorkee, Roorkee 247667, India}
 	\author{Vishnu Kumar Tiwari}
	\affiliation{Department of Physics, Indian Institute of Technology Kanpur, Kanpur 208016, India}
    \author{Yogendra Kumar}
	\affiliation{Research Institute for Synchrotron Radiation Science (HiSOR), Hiroshima University, Higashi-Hiroshima 739-0046, Japan}
 	\author{Vivek Kumar Malik}
	\affiliation{Department of Physics, Indian Institute of Technology Roorkee, Roorkee 247667, India}
    \author{Tulika Maitra}
	\affiliation{Department of Physics, Indian Institute of Technology Roorkee, Roorkee 247667, India}
    \author{Kiyohisa Tanaka}
	\affiliation{UVSOR-III Synchrotron, Institute for Molecular Science, Okazaki 444-8585, Japan}
    \author{Shinichiro Ideta}
	\affiliation{Research Institute for Synchrotron Radiation Science (HiSOR), Hiroshima University, Higashi-Hiroshima 739-0046, Japan}
    \author{Kenya Shimada}
	\affiliation{Research Institute for Synchrotron Radiation Science (HiSOR), Hiroshima University, Higashi-Hiroshima 739-0046, Japan}
	\affiliation{Research Institute for Semiconductor Engineering (RISE), Hiroshima University, Higashi-Hiroshima, 739-8527, Japan}
	\affiliation{International Institute for Sustainability with Knotted Chiral Meta Matter (SKCM$^2$), Hiroshima University, Higashi-Hiroshima, 739-8531, Japan}
    \author{Aftab Alam}
        	\email{aftab@iitb.ac.in}
	\affiliation{Department of Physics, Indian Institute of Technology Bombay, Mumbai 400076, India}
	\author{Jayita Nayak}
	\email{jnayak@iitk.ac.in}
	\affiliation{Department of Physics, Indian Institute of Technology Kanpur, Kanpur 208016, India}

\begin{abstract}
We have investigated the electronic structure and magnetotransport properties of SnSb$_6$Te$_{10}$ single crystals using density functional theory (DFT), synchrotron-based angle-resolved photoemission spectroscopy (ARPES), and quantum transport measurements. Our DFT calculations reveal a clear spin--orbit coupling driven band inversion between the Sb-$p$ and Te-$p$ states together with a non-trivial $\mathbb{Z}_2$ topological invariant. The calculated surface-state dispersion and hexagonally warped Fermi surface contours agree well with the ARPES measurements. Temperature-dependent transport measurements indicate dominant electron--phonon scattering, while Hall measurements confirm hole-type carriers with carrier density of the order of $10^{21}$~cm$^{-3}$. Both transverse and longitudinal magnetotransport exhibit weak antilocalization behavior, while Shubnikov-de Haas oscillations observed for $H \parallel c$ yield a Berry phase close to $\pi$, consistent with Dirac-like surface states. Furthermore, angle-dependent magnetotransport measurements reveal pronounced anisotropy associated with an anisotropic Fermi surface topology and mixed bulk-surface transport behavior. Our combined theoretical and experimental results establish SnSb$_6$Te$_{10}$ as a strong topological insulator and a promising platform for investigating topological transport phenomena in layered telluride systems.
\end{abstract}

     \maketitle

\section{Introduction}

Three-dimensional (3D) topological insulators (TIs) constitute a remarkable class of quantum materials characterized by insulating bulk states and symmetry-protected metallic surface states~\cite{TI1,TI2,TI3,TI4,kuroda2012experimental,Pb147,Mudgal_2024,ms813,PhysRevMaterials.9.044202}. These surface states originate from the interplay between strong spin--orbit coupling (SOC) and time-reversal symmetry, resulting in spin-momentum-locked Dirac fermions. Owing to their unusual electronic properties, TIs have attracted considerable interest in both fundamental condensed-matter physics and potential applications in spintronics, quantum information, microelectronics, and photonic devices. Experimentally, topological surface states can be directly visualized using angle-resolved photoemission spectroscopy (ARPES) and scanning tunneling spectroscopy (STS). In parallel, electronic transport measurements provide complementary information regarding the contribution of surface states to charge transport. In particular, Shubnikov-de Haas (SdH) oscillations~\cite{zhang2005,SdH,SDH2,Meena_2025,10.1063/5.0284677} and weak antilocalization (WAL)~\cite{PhysRevLett.106.166805,li2019quantitative,hikami1980spin} are widely regarded as important transport signatures of topological states arising from strong SOC and the associated non-trivial Berry phase.

Early studies on prototypical TIs such as Bi$_2$Se$_3$, Bi$_2$Te$_3$, and Sb$_2$Te$_3$~\cite{zhang2009topological} established the foundation for exploring more complex topological systems. Subsequently, several layered telluride compounds, including FeBi$_2$Te$_4$~\cite{saxena2020crystal}, MnBi$_2$Te$_4$~\cite{PhysRevB.99.155125,PhysRevB.100.155144}, SnBi$_2$Te$_4$~\cite{Eremeev2012,pan2015transport}, and PbBi$_2$Te$_4$~\cite{Pb124SdH,Pb124ARPES}, have emerged as promising platforms for investigating the interplay between topology, dimensionality, and magnetism. Among these, MnBi$_2$Te$_4$ and FeBi$_2$Te$_4$ belong to the class of magnetic TIs~\cite{deng2020quantum,chen2019intrinsic,guo2023novel}, whereas SnBi$_2$Te$_4$ and PbBi$_2$Te$_4$ are non-magnetic TIs with experimentally verified surface states~\cite{Eremeev2012,Pb124SdH,Pb124ARPES}. In particular, ARPES studies on the homologous series (SnBi$_2$Te$_4$)$_n$(Bi$_2$Te$_3$)$_m$ demonstrated that increasing Sn concentration suppresses the overlap between topological surface states and bulk conduction bands, making SnBi$_2$Te$_4$ especially favorable for realizing isolated topological surface states~\cite{PhysRevMaterials.5.014203}.

\begin{figure}
     \centering
     \includegraphics[width=0.45\textwidth]{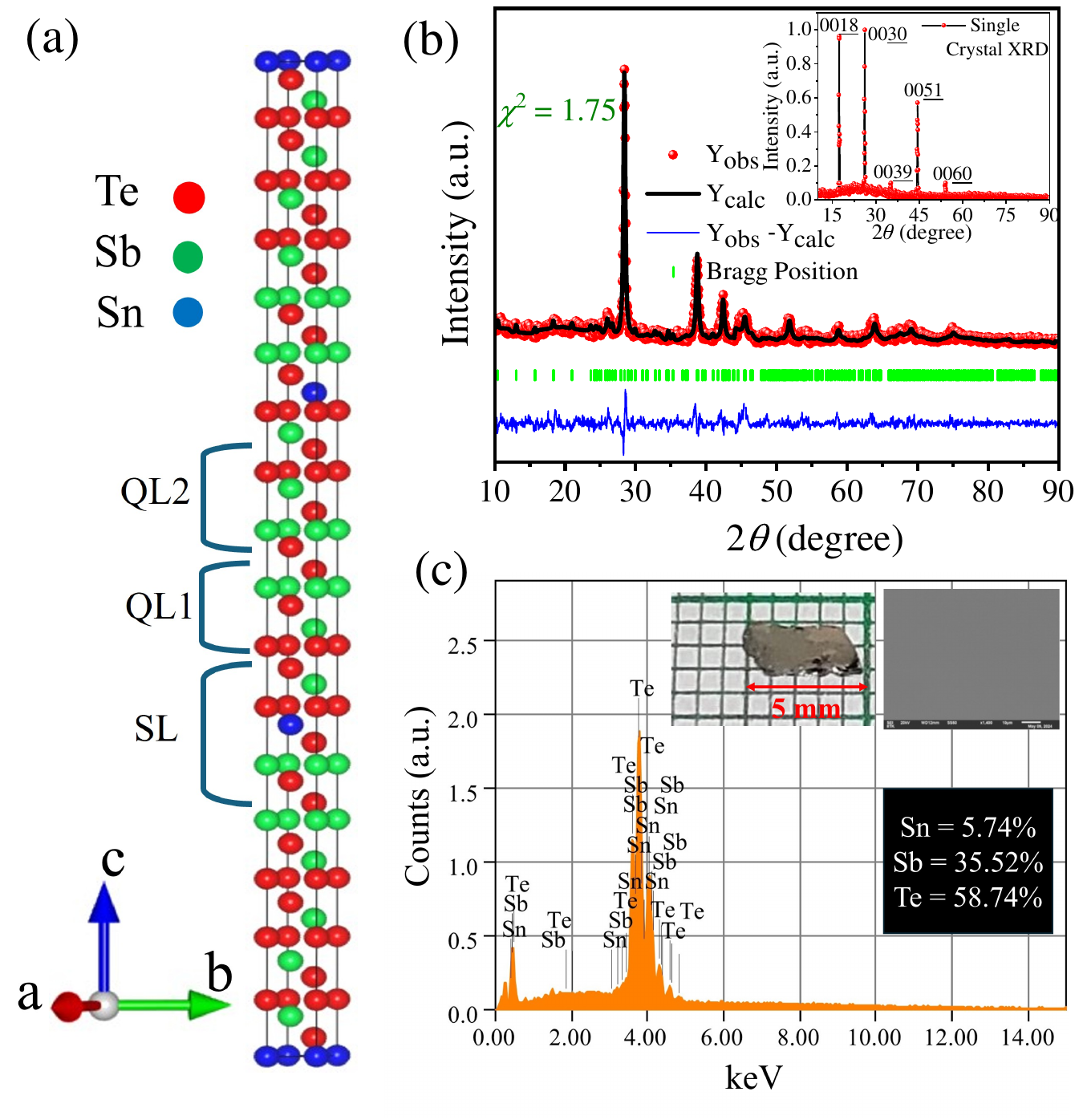}
     \caption{(a) Crystal structure of Sn-based ternary compound SnSb$_{6}$Te$_{10}$, with the ions of Sn, Sb, and Te shown in blue, green, and red colors, respectively. It is built with alternating SL and two QL blocks along the hexagonal $c$-axis. (b) Rietveld refinement of powder XRD pattern of SnSb$_{6}$Te$_{10}$ at 300 K. The inset shows the single crystal XRD of SnSb$_{6}$Te$_{10}$ at 300 K. (c) EDS spectrum over an area of SnSb$_{6}$Te$_{10}$ single crystal and SEM image (inset) and the atomic composition. A picture of the SnSb$_{6}$Te$_{10}$ crystal is shown in the inset with a typical size of 5 mm.
     }
     \label{fig:1}
\end{figure}

Another structurally related compound, SnSb$_2$Te$_4$~\cite{oeckler2011atom}, has also attracted attention due to its layered tetradymite-derived structure formed by the insertion of SnTe layers into the Sb$_2$Te$_3$ framework. Previous studies on SnSb$_2$Te$_4$ reported WAL behavior together with pressure-induced structural and electronic transitions~\cite{saxena2022structural}, indicating that Sn-based ternary tellurides provide a versatile platform for tuning topological electronic states.

First-principles calculations on the SnX$_2$Te$_4$, SnX$_4$Te$_7$, and SnX$_6$Te$_{10}$ (X = Sb, Bi) homologous series predicted several members of this family to host non-trivial topological phases~\cite{PhysRevB.92.045134}. In these compounds, alternating SLs and QLs stacked along the hexagonal axis strongly influence the electronic structure and spin texture. Interestingly, SnSb$_6$Te$_{10}$ was predicted to be a topologically trivial insulator, in contrast to most other members of the series~\cite{PhysRevB.92.045134}. However, the topological nature of SnSb$_6$Te$_{10}$ has remained experimentally unexplored.

In this work, we combine density functional theory (DFT), ARPES, and magnetotransport measurements to investigate the electronic structure and transport properties of SnSb$_6$Te$_{10}$ single crystals. Our experimentally refined crystal structure crystallizes in the centrosymmetric rhombohedral space group $R\overline{3}m$ (No.~166), and the corresponding electronic structure calculations reveal a clear SOC-driven band inversion between Sb-$p$ and Te-$p$ states. A subsequent Fu--Kane parity analysis yields a non-trivial $\mathbb{Z}_2$ invariant $(1;111)$, establishing SnSb$_6$Te$_{10}$ as a strong TI. The topological character is further supported by ARPES measurements, SdH oscillations with a Berry phase close to $\pi$, WAL behavior, and anisotropic magnetotransport consistent with hexagonally warped surface states. Our results, therefore, establish SnSb$_6$Te$_{10}$ as a promising platform for exploring topological surface transport in layered telluride systems.

\begin{figure*}[htb]
     \centering
     \includegraphics[width=0.9\linewidth]{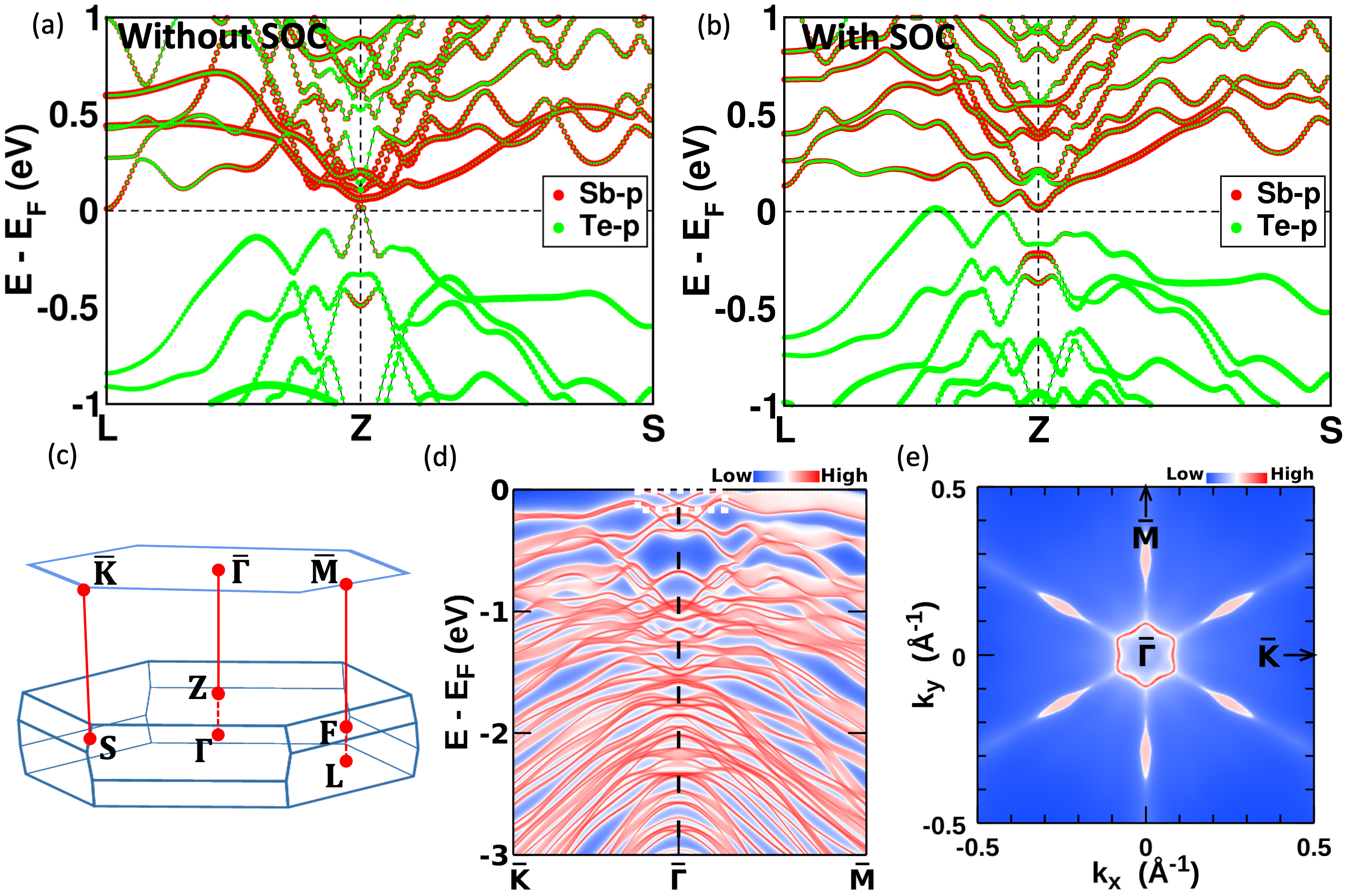}
     \caption{Electronic band structure of SnSb$_6$Te$_{10}$ along with the atomic orbital projections of Sb-$p$ (red color) and Te-$p$ (green color) atoms (a) without and (b) with SOC. (c) Bulk and surface BZ, marked with different high-symmetry points. (d) Simulated surface band dispersion and (e) Fermi surface for the conventional hexagonal (0001) surface of SnSb$_{6}$Te$_{10}$. The surface Dirac cone has been highlighted by a white dotted box in (d). The hexagonal warping of this surface Dirac cone is visible in (e) with projected bulk Fermi pockets along $\overline{\Gamma}$-$\overline{\text{M}}$.}
     \label{Fig1}
\end{figure*}

\section{Elemental analysis and crystal structure}

The details of single-crystal synthesis, structural characterization, stoichiometric analysis, and experimental methods are provided in Appendix~\ref{method}. Figure~\ref{fig:1}(a) shows the crystal structure of SnSb$_{6}$Te$_{10}$, where Sn, Sb, and Te atoms are represented by blue, green, and red spheres, respectively. The structure consists of alternating two quintuple layers (QLs) and one septuple layer (SL) stacked along the crystallographic $c$-axis. To assess phase purity, freshly cleaved single crystals were ground into fine powder for powder X-ray diffraction (XRD) analysis. The Rietveld refinement of the powder XRD pattern, shown in Figure~\ref{fig:1}(b), confirms that SnSb$_{6}$Te$_{10}$ crystallizes in the centrosymmetric rhombohedral space group $R\overline{3}m$ (No.~166). The refined lattice parameters are $a=b=4.2640$~\AA, $c=101.7036$~\AA, $\alpha=\beta=90^\circ$, and $\gamma=120^\circ$. The inset of Figure~\ref{fig:1}(b) displays the single-crystal XRD pattern, where only the $(00l)$ reflections are observed, confirming that the crystal growth direction is along the $c$-axis. \par
The elemental composition was examined using energy-dispersive X-ray spectroscopy (EDS) combined with tungsten scanning electron microscopy (W-SEM) on freshly cleaved crystal surfaces. No noticeable compositional variation was observed among different crystal pieces. Figure~\ref{fig:1}(c) presents the EDS spectrum together with the corresponding SEM image (inset) and the extracted atomic composition. The inset also shows an optical image of a typical SnSb$_{6}$Te$_{10}$ single crystal with a lateral size of approximately 5 mm.

\begin{figure}
    \centering
    \includegraphics[width=\columnwidth]{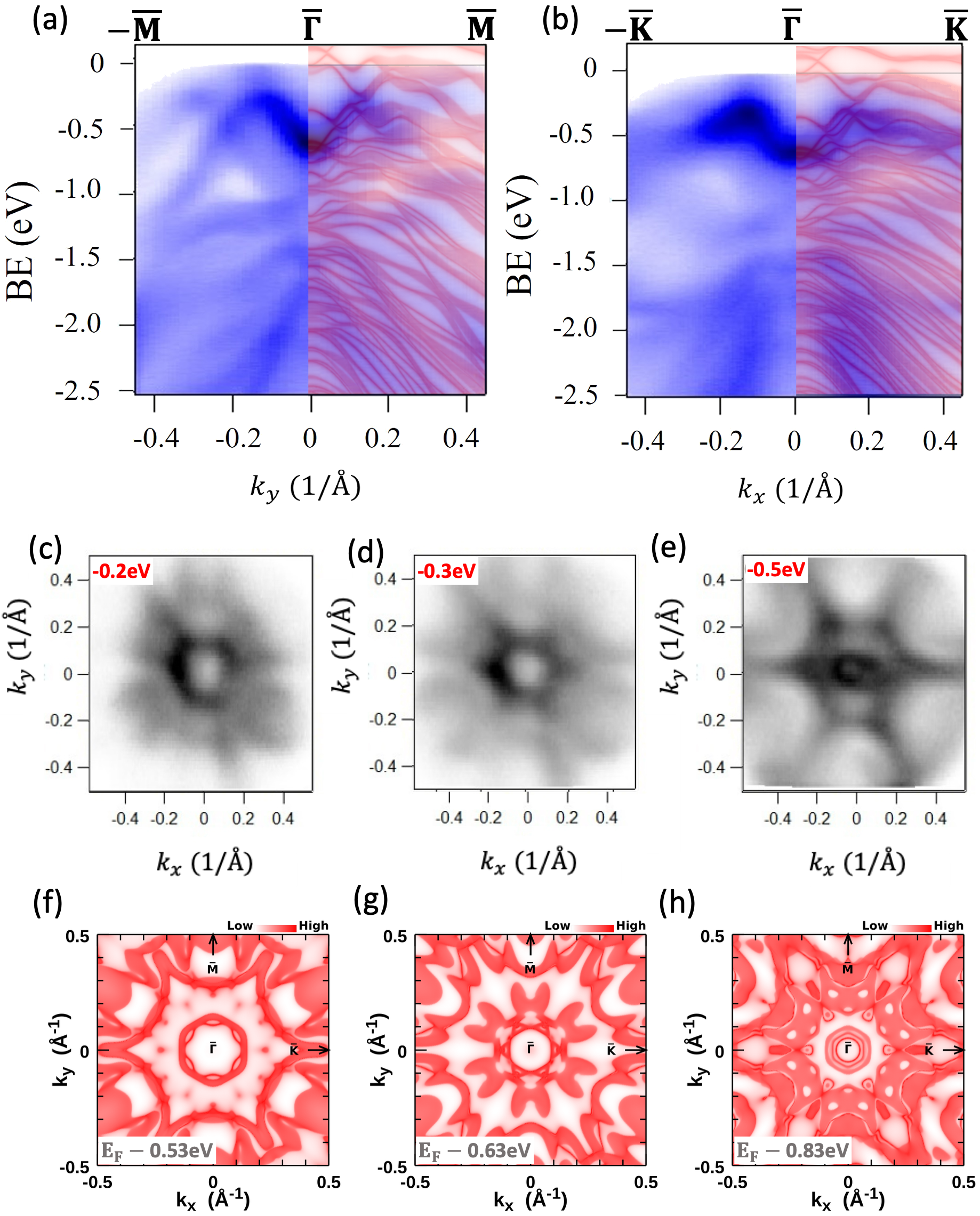}
    \caption{(a, b) Comparison of the experimental and energy-shifted simulated (0001) surface dispersion along -$\bar{\text{M}}-\bar{\Gamma}-\bar{\text{M}}$ and -$\bar{\text{K}}-\bar{\Gamma}-\bar{\text{K}}$. (c, d, e) The isoenergy surface maps of the ARPES spectra with incident photon energy h$\nu$ = 20 eV, at binding energies -0.2 eV, -0.3 eV, and -0.5 eV, respectively. (f, g, h) The corresponding simulated iso-energy contours, which are shifted by -0.33 eV below E$_F$ so as to match the observed and calculated chemical potentials.}
    \label{Fig3}
\end{figure}

\section{Electronic Structure}

\subsection{Bulk band structure and parity analysis}

Figures~\ref{Fig1}(a) and \ref{Fig1}(b) show the atom- and orbital-projected bulk band structures of SnSb$_6$Te$_{10}$ calculated without and with SOC, respectively. The computational details are provided in Appendix~\ref{comp}. In the presence of SOC, a clear band inversion between the Sb-$p$ and Te-$p$ states is observed near the Fermi level (E$_{\text{F}}$), indicating the non-trivial nature of the band topology. In addition, the valence band crosses the Fermi level along the L$-$Z direction, forming hole pockets consistent with the hole-dominated transport observed experimentally (see Sec.~\ref{Sec_WAL} and Fig.~\ref{fig:2''}). Figure~\ref{Fig1}(c) illustrates the bulk Brillouin zone (BZ) together with the projected (0001) hexagonal surface BZ and the corresponding high-symmetry points.
Although the calculated electronic structure exhibits finite spectral weight at the Fermi level, no direct band crossing occurs throughout the BZ, and a finite direct gap is preserved at every $k$-point. Therefore, the topological $\mathbb{Z}_2$ invariant remains well-defined throughout the BZ~\cite{WALalpha4,WAL32}.\par
To establish the topological character of SnSb$_6$Te$_{10}$, we evaluated the $\mathbb{Z}_2$ topological indices $(\nu_{0};\nu_{1}\,\nu_{2}\,\nu_{3})$ using the Fu--Kane parity criterion for centrosymmetric systems~\cite{WAL32}. The strong $\mathbb{Z}_2$ index is determined from:
\begin{equation}
(-1)^{\nu_0} = \prod^{8}_{i=1} \delta_i,
\label{z2}
\end{equation}
where, $\delta_i$ denotes the product of the parity eigenvalues of all occupied bands at the eight time-reversal invariant momenta (TRIM), namely the $\Gamma$ and Z points together with three equivalent L and F points. Our analysis shows that all TRIM points possess parity eigenvalue $-1$, except the Z point, which carries parity eigenvalue $+1$. The resulting parity inversion at the Z point yields the non-trivial topological invariant $(1;111)$, establishing SnSb$_6$Te$_{10}$ as a strong TI with robust topological surface states.

\subsection{Simulated Topological Surface States and ARPES Spectra}

Figure~\ref{Fig1}(d) shows the calculated surface band dispersion for the (0001) surface obtained by cleaving between the QL2 and SL (see Fig.~\ref{fig:1}(a)), where a Dirac-like surface state is observed around $\bar{\Gamma}$ in close proximity to the Fermi level (highlighted by a white dotted box). The corresponding surface Fermi contour is presented in Fig.~\ref{Fig1}(e) showing hexagonal warping of the surface Dirac cone. Experimentally, Hall measurements reveal a hole carrier density of the order of $10^{21}$~cm$^{-3}$ (see Fig.~\ref{fig:2''}(d)), which quantitatively matches with the calculated hole carrier density of 1.58$\times$$10^{21}$~cm$^{-3}$ at E$_{\text{F}}-$0.33 eV.
To account for this difference and enable a direct comparison with ARPES measurements, the calculated surface band structure was shifted  by 0.33~eV below Fermi level.\par
Figures~\ref{Fig3}(a) and \ref{Fig3}(b) compare the energy-shifted theoretical surface dispersion with the experimentally measured ARPES spectra along the -$\bar{\text{M}}-\bar{\Gamma}-\bar{\text{M}}$ and -$\bar{\text{K}}-\bar{\Gamma}-\bar{\text{K}}$ directions, respectively. A good overall agreement is obtained between theory and experiment. Although the Dirac point is not directly resolved in the ARPES spectra, the calculations place it slightly above the experimental Fermi level. As a result, the Dirac point overlaps with the conduction band region and remains experimentally inaccessible within the occupied states probed by ARPES.\par
Further evidence for the topological surface states is obtained from the evolution of the measured constant-energy contours shown in Figures~\ref{Fig3}(c--e). As the binding energy increases from $-0.2$ to $-0.5$~eV, the experimentally observed contours gradually evolve from nearly circular to strongly hexagonally warped shapes with pronounced sixfold symmetry. The corresponding simulated isoenergy contours, shown in Figures~\ref{Fig3}(f--h) after applying the same energy shift of $-0.33$~eV, reproduce this evolution remarkably well. The consistency between the calculated and measured surface-state contours, together with the SOC-driven band inversion and non-trivial $\mathbb{Z}_2$ invariant, provides strong evidence for the existence of topological surface states in SnSb$_6$Te$_{10}$.

\section{Resistivity}

\begin{figure}
     \centering
     \includegraphics[width=0.5\textwidth]{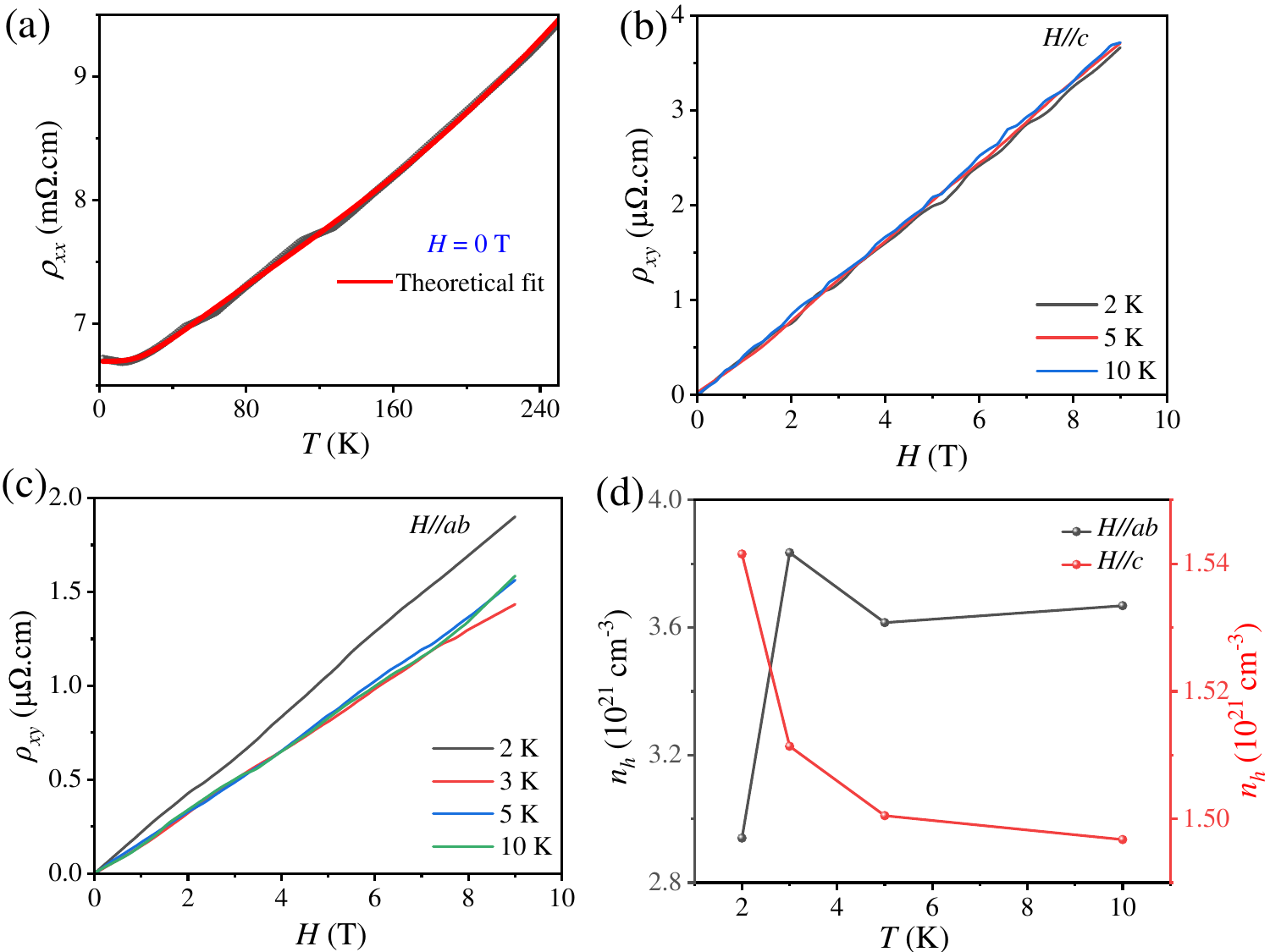}
     \caption{(a) Temperature dependence of longitudinal resistance of SnSb$_{6}$Te$_{10}$. The red line represents the fitting curve. (b, c) Hall resistivity at different temperatures for $H \parallel c$ and $H \parallel ab$, respectively. (d) Temperature-dependence of carrier density for $H \parallel c$ and $H \parallel ab$.}
     \label{fig:2''}
\end{figure}

Figure~\ref{fig:2''}(a) shows the temperature dependence of the longitudinal resistance $R_{xx}$ measured at zero magnetic field. The resistance decreases with decreasing temperature, indicating an overall metallic behavior, followed by a weak upturn below approximately 25~K. Such low-temperature upturns are commonly observed in several topological materials and may arise from enhanced surface-state contributions and quantum interference effects associated with WAL~\cite{ning2013one,Pb147device}. No clear anomaly corresponding to a structural or magnetic phase transition is observed over the measured temperature range.

The temperature dependence of $R_{xx}$ can be described using

\begin{equation}
R_{xx} = R_{0} + \beta e^{-\theta/T} + \gamma T^2,
\end{equation}

where $R_{0}$ denotes the residual resistance, the exponential term represents the contribution from electron-phonon scattering, and the $T^2$ term accounts for electron--electron interactions. The best fit yields $R_{0} = 6.6944$~$\Omega$, $\beta = 1.2601$~$\Omega$, $\theta = 87.87$~K, and $\gamma = 3.006 \times 10^{-5}$~$\Omega$~K$^{-2}$. The comparatively small value of $\gamma$ suggests that electron-phonon scattering dominates the transport behavior over the measured temperature range. Using the relation $\omega = k_B\theta/\hbar$, the corresponding characteristic phonon frequency is estimated to be approximately $1.15 \times 10^{13}$~rad/s.

\begin{figure*}
     \centering
     \includegraphics[width=0.95\textwidth]{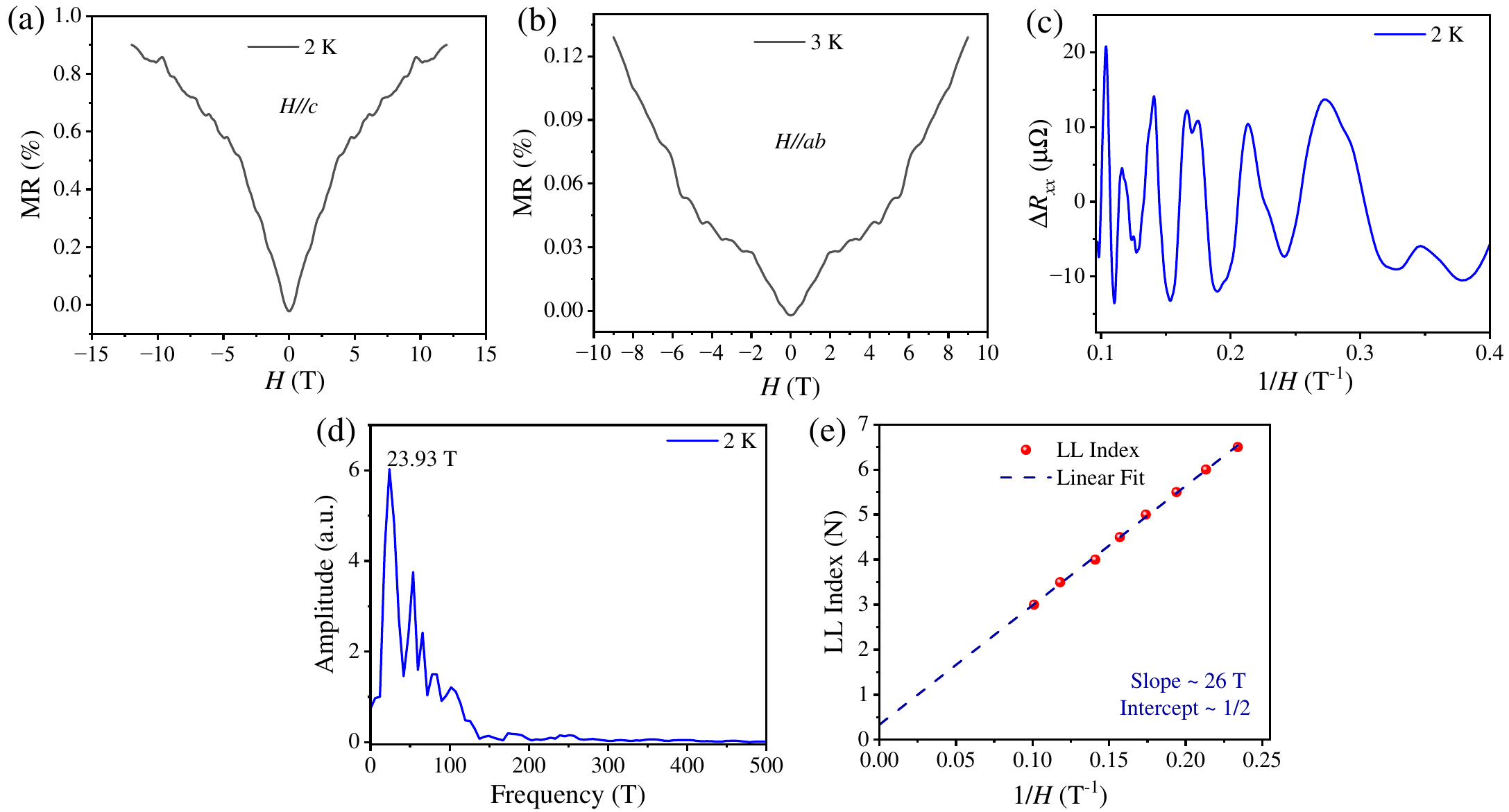}
     \caption{(a, b) TMR and LMR at 2 K temperature, respectively. (c) The SdH oscillations for the range of applied magnetic fields at 2 K for  $H \parallel c$. (d) FFT spectra of the oscillations at 2 K. (e) The linearly fitted Landau index diagram corresponds to the oscillations.}
     \label{fig:2'}
\end{figure*}

To further investigate the charge transport, Hall resistivity measurements were performed at different temperatures for both $H \parallel c$ and $H \parallel ab$ configurations, as shown in Figures~\ref{fig:2''}(b) and \ref{fig:2''}(c), respectively. The Hall resistivity was antisymmetrized using

\begin{equation}
\rho_{xy} = \frac{\rho_{xy}(H)-\rho_{xy}(-H)}{2},
\end{equation}

in order to eliminate the longitudinal magnetoresistance contribution arising from contact misalignment. In both field configurations, $\rho_{xy}$ exhibits a nearly linear field dependence with positive slope, indicating dominant hole-type charge carriers. The carrier density was extracted from the linear Hall coefficient using $R_H = 1/ne$.

The estimated hole carrier density $n_h$ is of the order of $10^{21}$~cm$^{-3}$, comparable to previously reported values for related topological materials~\cite{Bi2Se32012quantized,Pb124SdH,ms813}. The temperature dependence of $n_h$ is shown in Figure~\ref{fig:2''}(d). While $n_h$ decreases gradually with increasing temperature for $H \parallel c$, an opposite trend is observed for $H \parallel ab$, indicating anisotropic transport behavior consistent with the layered crystal structure and anisotropic Fermi surface topology.

\section{Magnetoresistance and SdH Oscillations}

Figures~\ref{fig:2'}(a) and \ref{fig:2'}(b) show the magnetic-field dependence of the transverse magnetoresistance (TMR) and longitudinal magnetoresistance (LMR) of SnSb$_6$Te$_{10}$ measured at 2~K, respectively, with the magnetic field applied perpendicular to the basal plane of the crystal. The magnetoresistance is defined as

\begin{equation}
\mathrm{MR} = \frac{\rho_{xx}(H)-\rho_{xx}(0)}{\rho_{xx}(0)}.
\end{equation}

To eliminate the contribution arising from Hall-voltage mixing due to contact misalignment, the longitudinal resistivity data were symmetrized using

\begin{equation}
\rho_{xx}(H)=\frac{\rho_{xx}(H)+\rho_{xx}(-H)}{2}.
\end{equation}

Both TMR and LMR exhibit a sharp low-field cusp characteristic of WAL, which will be discussed in the following section. In addition, clear SdH oscillations are observed in the high-field region of the TMR data [Fig.~\ref{fig:2'}(c)], consistent with previous reports on topological materials~\cite{Bi2Te3sdh,Pb124SdH,Pb147device}.

The SdH oscillations arise from the quantization of cyclotron orbits into Landau levels in a strong magnetic field. To extract the oscillatory component $\Delta R_{xx}$, a smooth polynomial background was subtracted from the magnetoresistance data~\cite{smoothBG2017}. The resulting oscillatory signal is shown in Figure~\ref{fig:2'}(c), where the oscillation amplitude decreases gradually with increasing $1/H$, as expected for quantum oscillations. The fast Fourier transform (FFT) spectrum [Fig.~\ref{fig:2'}(d)] reveals a dominant oscillation frequency of 23.93~T together with several weaker frequencies of smaller amplitude.\par
The Landau level fan diagram corresponding to the SdH oscillations is shown in Figure~\ref{fig:2'}(e). Integer Landau indices were assigned to the maxima of $\Delta R_{xx}$~\cite{LKequscience}. The linear extrapolation of the Landau index plot yields an intercept close to 1/2, suggesting a non-trivial Berry phase associated with Dirac-like charge carriers. The oscillation frequency extracted from the slope of the linear fit is consistent with the dominant FFT frequency. Using the relation $\gamma=\phi_B/2\pi$, the extracted intercept corresponds to a Berry phase close to $\pi$, supporting the existence of topologically non-trivial surface states in SnSb$_6$Te$_{10}$.

Further, the Fermi wave vector was estimated using Onsager's relation~\cite{Fermiwavevector},

\begin{equation}
f_{\mathrm{SdH}}=\left(\frac{\hbar}{2\pi e}\right)\pi k_F^2,
\end{equation}

which yields $k_F \approx 0.027$~\AA$^{-1}$ for the dominant oscillation frequency of 23.93~T.

\begin{figure}
     \centering
     \includegraphics[width=0.480\textwidth]{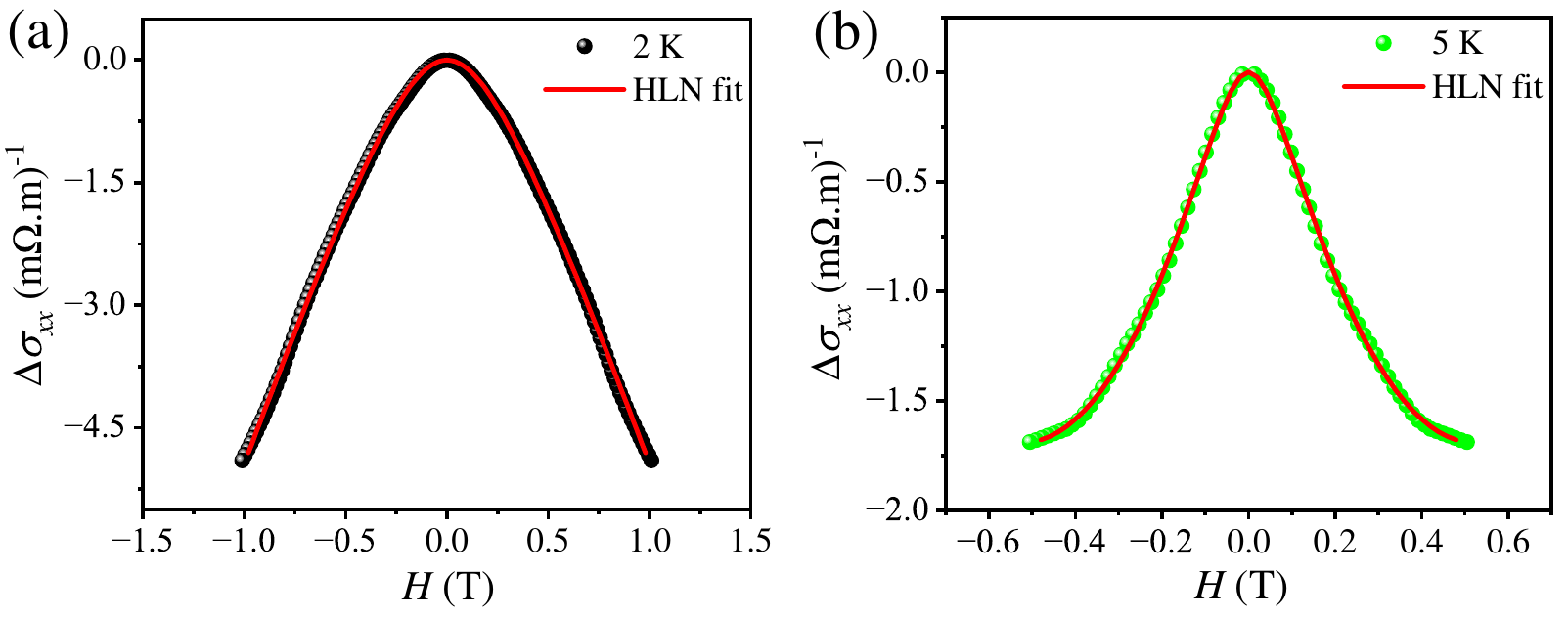}
     \caption{(a, b) Magnetoconductance ($\Delta\sigma_{xx}$) fitted using the HLN model at 2 K and 5 K for $H \parallel c$, respectively.}
     \label{fig:2}
\end{figure}

\section{Weak Antilocalization} \label{Sec_WAL}

WAL originates from strong SOC and the associated $\pi$ Berry phase acquired by charge carriers undergoing time-reversed scattering processes. In topological materials, the destructive quantum interference between these time-reversed electron trajectories suppresses backscattering and gives rise to enhanced conductivity near zero magnetic field~\cite{WAL32,WAL29}. Such WAL behavior is commonly observed in systems with strong SOC and topological electronic states.

Figures~\ref{fig:2}(a) and \ref{fig:2}(b) show the magnetic-field dependence of the magnetoconductance $\Delta \sigma_{xx}$ measured at 2~K and 5~K for $H \parallel c$, respectively. Here,

\begin{equation}
\Delta \sigma_{xx} = \sigma_{xx}(H)-\sigma_{xx}(0),
\end{equation}

where $\sigma_{xx}=\rho_{xx}/(\rho_{xx})^2+(\rho_{xy})^2$. The pronounced low-field cusp together with the negative magnetoconductance is characteristic of WAL behavior~\cite{WAL29,wal30,Pb124SdH}. The overall field dependence of $\Delta \sigma_{xx}$ further suggests the coexistence of both surface and bulk transport channels.

To analyze the WAL response for $H \parallel c$, the low-field magnetoconductance was fitted using the modified Hikami--Larkin--Nagaoka (HLN) model~\cite{wal33},

\begin{figure*}
     \centering
     \includegraphics[width=0.95\textwidth]{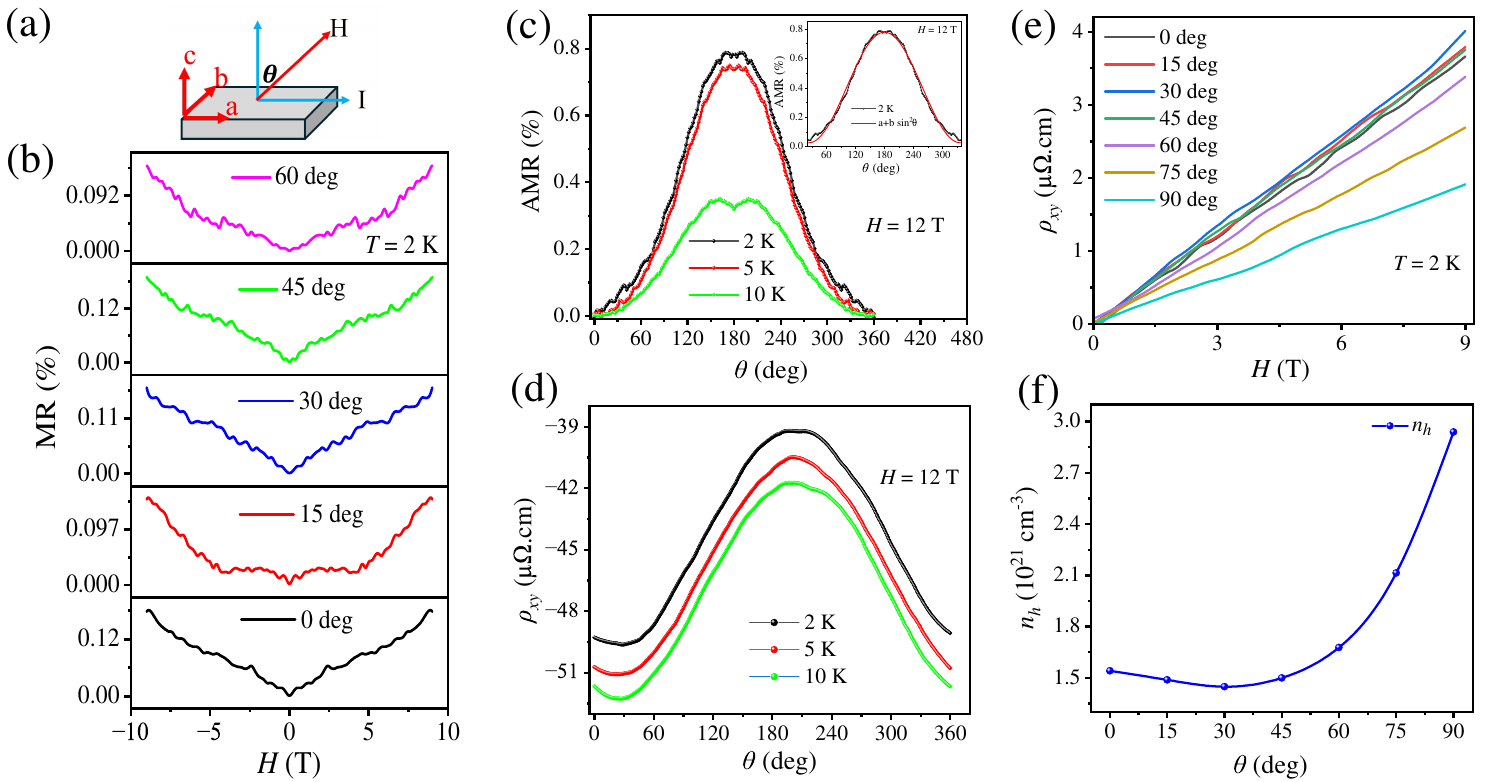}
     \caption{(a) The schematic illustration of the directional setup for all angle-dependent measurements. (b) MR at different angles to the magnetic field at 2 K. (c) AMR at various temperatures and at a 12 T magnetic field. The inset shows the $sin^2\theta$ fitting of AMR at 2 K. (d) Angle-dependent Hall resistivity at various temperatures and at a 12 T magnetic field. (e) The Hall resistivity at different angles to the magnetic field at 2 K. (f) The variation of hole carrier density with the angle to the magnetic field.}
     \label{fig:3}
\end{figure*}

\begin{equation}
\Delta \sigma_{xx}(H) = -C \left[\Psi \left(\frac{1}{2}+\frac{h}{4eHl_{\phi}^{2}}\right)-\ln \left(\frac{h}{4eHl_{\phi}^{2}}\right)\right],
\end{equation}

where $C=\alpha e^2/\pi h$, $\Psi$ is the digamma function, $\alpha$ is the fitting coefficient associated with the effective conduction channels, and $l_{\phi}$ denotes the phase coherence length. The modified HLN expression incorporates additional contributions arising from bulk transport and spin--orbit scattering at higher magnetic fields.\par
The fitting results are shown in Figures~\ref{fig:2}(a) and \ref{fig:2}(b). The extracted phase coherence length decreases from 58.92~nm at 2~K to 27.49~nm at 5~K, indicating enhanced inelastic scattering and reduced phase coherence at elevated temperatures. Since the estimated coherence length remains significantly smaller than the crystal thickness, the observed WAL response is consistent with predominantly 3D transport. The HLN analysis is used here to characterize the WAL behavior and to estimate the phase-coherence length.

\section{Angle-dependent measurements}

To further investigate the anisotropic transport behavior of SnSb$_6$Te$_{10}$, angle-dependent magnetoresistance (AMR) and Hall resistivity measurements were performed. The inset of Figure~\ref{fig:3}(a) illustrates the measurement geometry. In this configuration, the magnetic field is applied perpendicular to the crystal's basal plane while the sample is rotated about the $b$-axis, allowing the field orientation to evolve continuously from in-plane to out-of-plane.

Figure~\ref{fig:3}(b) shows the magnetoresistance measured at 2~K under an applied magnetic field of 12~T for different field orientations. The MR exhibits an overall parabolic field dependence together with a pronounced low-field WAL cusp that is strongest near $\theta = 0^\circ$ and becomes significantly suppressed around $\theta = 60^\circ$. A weak WAL feature reappears near $\theta = 90^\circ$. The strong angular dependence of the WAL response suggests anisotropic quantum transport, with possible contributions from surface states.\par
The angular dependence of the AMR measured at different temperatures under 12~T is presented in Figure~\ref{fig:3}(c). The AMR exhibits a twofold symmetric behavior with maxima near $\theta = 180^\circ$ and minima around $\theta \approx 0^\circ$ and $360^\circ$, indicating pronounced transport anisotropy. The AMR amplitude decreases by nearly 50\% at 10~K compared to 2~K, suggesting a gradual enhancement of bulk conduction at elevated temperatures. The inset of Figure~\ref{fig:3}(c) shows that the angular dependence can be described by the relation $a+b\sin^2\theta$. Such $\sin^2\theta$ behavior is consistent with an anisotropic Fermi surface topology and is commonly observed in layered topological materials with hexagonally warped electronic states~\cite{PhysRevLett.108.216803}. Figure~\ref{fig:3}(d) presents the angular dependence of the Hall resistivity measured at different temperatures. The Hall resistivity exhibits clear angular modulation with a maximum near $200^\circ$ and a minimum around $30^\circ$. Figure~\ref{fig:3}(e) shows the field-dependent Hall resistivity at different angular orientations. In all configurations, the positive Hall slope confirms dominant hole-type charge transport. The corresponding angular variation of the hole carrier density is shown in Figure~\ref{fig:3}(f), where the carrier density increases gradually as the field orientation changes from $\theta = 0^\circ$ to $\theta = 90^\circ$. The observed anisotropic Hall response further supports the presence of an anisotropic electronic structure in SnSb$_6$Te$_{10}$.

\section{Conclusion}

In summary, we have investigated the electronic structure and magnetotransport properties of SnSb$_6$Te$_{10}$ single crystals using a combination of DFT, ARPES, and quantum transport measurements. The electronic structure calculations reveal a clear SOC-driven band inversion together with a non-trivial $\mathbb{Z}_2$ topological invariant, establishing SnSb$_6$Te$_{10}$ as a strong TI. The calculated surface-state dispersion and isoenergy contours agree well with the ARPES measurements, including the evolution of hexagonally warped constant-energy contours.
Transport measurements further support the topological nature of the system. The temperature-dependent resistivity indicates dominant electron--phonon scattering, while Hall measurements confirm hole-type charge carriers with anisotropic transport behavior. SdH oscillations yield a Berry phase close to $\pi$, consistent with Dirac-like surface states. In addition, observations of WAL behavior and anisotropic magnetotransport suggest the coexistence of bulk and surface transport contributions.

Overall, our combined theoretical and experimental results establish SnSb$_6$Te$_{10}$ as a promising layered topological material hosting non-trivial surface states and anisotropic electronic transport. These findings provide a foundation for further investigations of topological transport phenomena in Sn-based layered telluride systems.

\begin{acknowledgements}
    
The authors acknowledge IIT Kanpur and the Science and Engineering Research Board, India (project no. CRG/2023/00786060, and SPG/2021/000443) for financial support. The authors acknowledge the Research Institute for Synchrotron Radiation Science (HiSOR), Hiroshima University (proposal no. 23AG039), and UVSOR-III Synchrotron, Institute of Molecular Science (proposal nos. 25IMS6660 and 25IMS6677), Japan, for ARPES measurements. SM and BD thank the Department of Physics, IIT Bombay, for HPC support and the Ministry of Education, Govt. of India, for financial support.
\end{acknowledgements}

\section*{Data Availability Statement}
The datasets generated during and/or analysed during the current study are available from the corresponding author on reasonable request.

\appendix

\section{Experimental Details and Methods}\label{method}

Single crystals of SnSb$_6$Te$_{10}$ were synthesized using the flux-growth technique. High-purity Sn (powder, 99.95\%, Alfa Aesar), Sb (shots, 99.999\%, Alfa Aesar), and Te (lumps, 99.999\%, Alfa Aesar) were mixed in the molar ratio 1:10:16 and sealed in an evacuated quartz tube under high vacuum. The sealed ampoule was heated to 800~$^\circ$C and maintained at this temperature for 15~h to ensure homogeneous melting. Subsequently, the melt was cooled slowly to 595~$^\circ$C at a rate of 10~$^\circ$C/h, followed by further cooling to 582~$^\circ$C at 1~$^\circ$C/h. The excess flux was removed by centrifugation at 582~$^\circ$C, yielding plate-like single crystals of SnSb$_6$Te$_{10}$.
The obtained crystals typically possessed lateral dimensions of approximately $3.0 \times 3.0$~mm$^2$, with the naturally cleaved surface corresponding to the crystallographic $ab$ plane. The crystal structure was characterized using XRD on a PANalytical X'Pert PRO diffractometer with Cu K$\alpha_1$ radiation. The elemental composition was verified using energy-dispersive X-ray spectroscopy (EDS) performed on a JEOL JSM-6010LA scanning electron microscope.
Electrical and magnetotransport measurements were performed using a standard four-probe configuration in a 12~T Physical Property Measurement System (PPMS, Quantum Design). Electrical contacts were made using thin platinum wires attached with conductive silver paste.
Angle-resolved photoemission spectroscopy (ARPES) measurements using synchrotron radiation were performed at beamline BL-1 of the Research Institute for Synchrotron Radiation Science (HiSOR), Hiroshima University, and at beamline BL5U of the UVSOR-III Synchrotron, Institute for Molecular Science, Japan~\cite{SHIMADA2001504,Iwasawa:rx5035}. The energy and angular resolutions were set to approximately 20~meV and 0.1$^\circ$, respectively. The samples were cleaved \textit{in situ} at approximately 20~K using the ceramic top-post method under an ultrahigh vacuum better than $3.0 \times 10^{-9}$~Pa.

\section{Computational Details}\label{comp}

To investigate the electronic structure of SnSb$_6$Te$_{10}$, first principles calculations were performed within the framework of Kohn-Sham DFT \cite{DFT1,DFT2} using the Vienna \textit{Ab-initio} Simulation Package (VASP) \cite{VASP1,VASP2,VASP3} based on the projector-augmented wave (PAW) \cite{PAW1,PAW2} method. The Perdew--Burke--Ernzerhof (PBE) \cite{PBE} functional within the generalized gradient approximation (GGA) was employed to describe the exchange-correlation interactions. Long-ranged van der Waal's interaction was included through the DFT-D3 dispersion correction method by Grimme \textit{et al.} \cite{VdW} as implemented in VASP. The first BZ was sampled using a $\Gamma$-centered $9 \times 9 \times 9$ $k$-mesh and kinetic energy cutoff fixed at 340 eV, with the self-consistent electronic energy convergence criterion set to $10^{-6}$~eV. For band structure calculations, 100 $k$-point divisions were chosen along each high-symmetry path. A modified version of the \textsc{WaveTrans} \cite{Wavetrans} code was used to read the electronic wavefunctions and perform a parity analysis \cite{WAL32}, enabling the determination of the topological $\mathbb{Z}_2$ invariants \cite{WALalpha4}.
Maximally localized Wannier functions (MLWFs) \cite{MLWF1,MLWF2,MLWF3} were generated using the \textsc{Wannier90} \cite{w90_1,w90_2,w90_3} package, employing projections onto the $s$ and $p$ orbitals of Sn, Sb, and Te atoms to construct a tight-binding Hamiltonian from pre-converged DFT results. The surface dispersion, Fermi surface and isoenergy contours were then obtained by applying the iterative Green's function method \cite{green1,green2,green3,green4} on a semi-infinite slab, as implemented in the \textsc{WannierTools} \cite{Wtools} package.

\bigskip

\bibliography{reference}

\end{document}